\documentclass[twocolumn,
nofootinbib,
 amsmath,amssymb,
 aps,
 prd,
 floatfix,
]{revtex4-2}
\usepackage{subcaption}
\usepackage{graphicx}
\usepackage{dcolumn}
\usepackage{bm}
\usepackage{slashed}
\usepackage[normalem]{ulem}
\usepackage[export]{adjustbox}
\usepackage{xcolor}

\newcommand{\appropto}{\mathrel{\vcenter{
  \offinterlineskip\halign{\hfil$##$\cr
    \propto\cr\noalign{\kern2pt}\sim\cr\noalign{\kern-2pt}}}}}

\usepackage{hyperref}

\newcommand{\vEDM}{\vec E_\mathrm{DM}}

\begin{document}



\title{Improved modelling for dark photon detection with dish antennas}%



\author{Jordan Gu\'e}
\email{jordan.gue@obspm.fr}
\affiliation{SYRTE, Observatoire de Paris, Universit\'e PSL, CNRS, Sorbonne Universit\'e, LNE, 61 avenue de l’Observatoire 75014 Paris, France\\
 }%
\author{Aur\'elien Hees}
\affiliation{SYRTE, Observatoire de Paris, Universit\'e PSL, CNRS, Sorbonne Universit\'e, LNE, 61 avenue de l’Observatoire 75014 Paris, France\\
}%
\author{Etienne Savalle}
\affiliation{IRFU, CEA, Universit\'e Paris-Saclay, F-91191 Gif-sur-Yvette, France\\}
\author{Laurent Chevalier}
\affiliation{IRFU, CEA, Universit\'e Paris-Saclay, F-91191 Gif-sur-Yvette, France\\}
\author{Pierre Brun}
\affiliation{IRFU, CEA, Universit\'e Paris-Saclay, F-91191 Gif-sur-Yvette, France\\}
\author{Peter Wolf}
\affiliation{SYRTE, Observatoire de Paris, Universit\'e PSL, CNRS, Sorbonne Universit\'e, LNE, 61 avenue de l’Observatoire 75014 Paris, France\\
}%

\date{\today}

\begin{abstract}

A vector dark matter candidate, also known as dark photon, would induce an oscillating electric field through kinetic mixing. One detection strategy uses a spherical reflector to focus the induced emission at its center of curvature. On one hand, we investigate the effects of diffraction in this type of experiment from an analytical standpoint, making use of the Kirchhoff integral theorem in the low-curvature dish limit. On the other hand, we estimate the impact of mode-matching, in the case of detection by a pyramidal horn antenna. We show that the expected signal intensity can be significantly reduced compared to usual estimates. Our method is applied to the re-interpretation of the SHUKET experiment data, the results of which are shown to be degraded by a factor of $\sim$~50 due to both diffraction and mode-matching. The analytical method allows optimizing some experimental parameters to gain sensitivity in future runs. Our results can be applied to any dish antenna experiment using a low curvature reflector.

\end{abstract}

\maketitle


\section{\label{sec:Intro}Introduction}

While required to explain several astrophysical and cosmological observations, the microscopic nature of dark matter (DM) is still to this day one of the biggest mysteries in physics \cite{bertone:2018aa}. Among many other classes of DM, ultra-light dark matter (ULDM) models have recently gained a lot of attention in the scientific community, due to the absence of signals from historically dominant models, such as WIMPs. These models are characterized by particles with low mass, from $10^{-22}$ eV to $0.1$ eV, meaning that their detection in particle colliders is complicated. Thus, other kind of experiments have to be considered for their detection.

One particularly well motivated model of ULDM is called the dark photon (DP), a massive spin-1 field (of mass $m$) which appears in many Beyond the Standard Model theories. The DP field is initially frozen after inflation and starts to oscillate at its Compton frequency when $mc^2/\hbar \gg H$ (where $c$ is the speed of light, $\hbar$ the reduced Planck constant and $H$ the Hubble parameter), and it can be shown that these oscillations scale as $a^{-3}(t)$ (where $a$ is the cosmological scale factor), at late cosmological times, behaving as cold dark matter (CDM). The production of CDM in this case is done through non-thermal processes, such as the so called misalignment mechanism \cite{nelson:2011tv,Arias}. This additional $U(1)$ field is also well motivated on the particle physics side, as the DP model is a particular case of the so-called U boson, which makes the minimal gauged Standard Model extension, through its coupling with the B-L current of the Standard Model \cite{U-boson}.

For all these reasons, the DP is an interesting light dark matter candidate, and experiments hunting for it are arising more and more. In the following, we will be interested in the coupling of the DP field with the electromagnetic (EM) field. The strength of this coupling is parameterised by the dimensionless parameter $\chi$. This coupling induces the appearance of an electric field filling the whole space, with amplitude directly proportional to $\chi$ \cite{Arias, Horns}. Numerous experiments, e.g \cite{SHUKET,Tokyo1,Tokyo2,Tokyo3,Tokyo4,Qualiphide, Kotaka23, Knirck23, Bajjali23}, use a parabolic or spherical mirror to focus this small electric field in order to enhance it inside a detector located at the curvature center of the mirror. 
In the initial proposal of such kind of experiments, \cite{Horns} made the assumption that the power generated by the dish is entirely focused on the center of curvature, as long as diffraction effects are negligible, i.e when the Compton wavelength of the DP field $\lambda = 2\pi \hbar/mc$ is much larger than the radius of the mirror $r$. In many of the experiments using such setups, numerical simulations were performed to estimate the loss through diffraction effects. However, no analytical derivation of such losses has been done in our knowledge.

In this paper, we derive an analytical expression of the EM power measured by the antenna detector as a function of the DP Compton frequency and of the experimental parameters. This calculation is performed  in two steps: (i) we compute the electric field at the location of the antenna and (ii) we determine how much of this electric power is measured by the antenna. Regarding the first point, we provide an analytical derivation for the electric field by propagating the electric field generated by the dish using Kirchhoff integral theorem. This analytical expression is obtained by using some assumptions on the setup, in particular considering small curvature, i.e the typical size of the mirror being much smaller than the curvature radius (see Fig.~\ref{fig:SHUKET_Kirchhoff} for the definition of the dish radius and of its curvature radius). In particular, we show that diffraction effects can significantly impact  the electric field at the location of the detector. Second, we turn to the detection of the electric field and focus on horn antenna detectors. We demonstrate how to derive analytically the EM power measured by the detector using overlap integral and show that it can impact significantly the measurements. Considering both effects, we derive an efficiency coefficient, i.e the ratio between the power measured by the horn antenna and the power emitted by the dish. 

As a practical illustration, we consider a setup used in the SHUKET experiment \cite{SHUKET}, to show how both diffraction of the field and overlap integral of modes affect the sensitivity of the experiment. We show that the efficiency coefficient in such case is very small, of the order of $10^{-4}$, leading to a loss of sensitivity of SHUKET to the coupling $\chi$ by a factor $\sim 50$. We show that diffraction effects and overlap integral of modes contribute almost equally to this small efficiency coefficient, implying that both effects must be taken into account very carefully. Finally, we show that using the analytical expressions derived in the paper, one can carefully optimize the experimental parameters of the experiment in order to maximize the signal amplitude. In the case of SHUKET, we show that modifying the dish-antenna detector distance and DP frequency of interest would lead to a reduced loss on the power received by a factor $\sim 9$ compared to the initial set of parameters. This work could also be used in the readjustment of the sensitivity of other already existing experiments using similar setups, e.g. \cite{Qualiphide,Bajjali23, Knirck23}. In addition, it could be useful for the modelling of future experiments aiming at detecting EM waves of comparable frequencies using parabolic or spherical antennas as EM emitter. 

The paper is organised as follows : we first recall respectively the form of the electric field induced by the photon-DP coupling and the total power emitted by a spherical dish immersed in this field in Sec.~\ref{sec:pheno_DP} and Sec.~\ref{sec:emission_field}. Then, in Sec.~\ref{sec:propag_field}, we derive an analytical expression of the field propagating from the dish emitter. In Sec.~\ref{sec:detection_field}, we turn to the detection of the field using a horn antenna and we derive an analytical expression of the ratio between the received power by the antenna and the emitted power from the dish, using overlap integral of modes on one hand and antenna factor on the other hand. Finally, in Sec.~\ref{sec:SHUKET}, we use all the previous results to derive the expected power received by the horn antenna with the parameters of the SHUKET experiment \cite{SHUKET} and we propose another set of parameters, involving e.g. the DP frequency of interest, which would minimize the loss of sensitivity of the experiment through Kirchhoff integral and overlap integral of modes. 

\section{\label{sec:pheno_DP}Dark photon field induced electromagnetic field}

The Lagrangian describing the interaction between the EM field 4-potential $A^\mu$ and a DP field 4-potential $\phi^\mu$ of mass $m$ is given by \cite{Horns}\footnote{The unit of the DP 4-potential  field $\phi^\mu$ is V.s.m$^{-1}$, as the usual EM 4-potential $A^\mu$.}
\begin{align}\label{lagrangian}
  \mathcal L =& - \frac{1}{4\mu_0}F^{\mu\nu}F_{\mu\nu}+j^\mu A_\mu  - \frac{1}{4\mu_0}\Phi^{\mu\nu}\Phi_{\mu\nu} \\ \nonumber
  &\qquad - \frac{m^2c^2}{2\mu_0\hbar^2}\phi^\mu\phi_\mu -\frac{\chi}{2\mu_0}F_{\mu\nu}\Phi^{\mu\nu}\, ,
\end{align}
where $F_{\mu\nu}=\partial_\mu A_\nu - \partial_\nu A_\mu$ is the usual electromagnetic field strength tensor, $\Phi_{\mu\nu}=\partial_\mu \phi_\nu - \partial_\nu \phi_\mu$ is the DP field strength tensor, $\chi$ is the kinetic mixing coupling parameter which characterizes the coupling between the DP and the EM field and $j^\mu$ is the usual matter electromagnetic 4-current. 

Solving the field equation of the DP field in vacuum, one finds the solution 
\begin{subequations}
\begin{align}
    \phi^\alpha= Y^\alpha e^{ik_\mu x^\mu} \, ,
\end{align}
with 
\begin{align}
    k_\mu k^\mu = -\frac{\omega^2}{c^2} + \left|\vec k\right|^2 = -\left(\frac{mc}{\hbar}\right)^2 \, .
\end{align}
\end{subequations} 
Then, one can show that the DP-EM coupling induces an ordinary electromagnetic field in vacuum (see e.g. \cite{Horns,Gue23} for more details)
\begin{equation}\label{eq:A}
	A^\beta = -\chi \phi^\beta = -\chi Y^\beta e^{i k_\mu x^\mu}\, . 
\end{equation}

The amplitude of oscillation of the DP field, $\left|\vec Y\right|$, depends directly on the local DM energy density $\rho_{\mathrm{DM}}$ as
\begin{equation}\label{energy_density}
  \rho_\mathrm{DM} = \frac{\omega^2 |\vec Y|^2}{2\mu_0c^2}\, ,
\end{equation} 
for $\vec k=0$. Using recent data, its experimental value is estimated to be $\rho_\mathrm{DM}=\left(0.55 \pm 0.17\right)$ GeV/cm$^3$\cite{GaiaSausage}.

Considering an EM emitter of size $r \sim \mathcal{O}$(m) (see Sec.~\ref{sec:SHUKET}) and DM frequency $f \sim \mathcal{O}$(GHz), one can approximate the DM field as a standing wave, i.e with a negligible propagating component. Indeed, the wave vector of the massive DP field is $\vec k\approx \omega \vec v/c$ and therefore $\vec k \cdot \vec x \leq |\vec k| r = 2 \pi f r v_\mathrm{DM}/c^2 \ll 1$, where $v_\mathrm{DM} \sim 10^{-3}c$ is the typical galactic velocity in the DM halo. Then, the induced electromagnetic field consists only in an oscillating electric field \cite{Horns}
\begin{subequations}\label{eqs:EDM}
\begin{equation}
	 E^j_\mathrm{DM} = -\frac{\partial  A^j}{\partial t} = \Re\left[-i \chi \omega  Y^j e^{-i \omega t}\right]\, .
\end{equation}
The amplitude of this oscillating electric field is therefore directly related to the local DM density through
\begin{equation}
	\left|\vEDM\right| = \chi c\sqrt{2\mu_0 \rho_{\mathrm{DM}}}\, .
 \label{amp_E_field}
\end{equation}
\end{subequations}

The idea of several experiments searching for DP is to focus this possible small electromagnetic field present everywhere in space and to enhance it by using reflecting materials in order to hopefully make it detectable, e.g. \cite{SHUKET,Tokyo1,Tokyo2,WISPDMX,Gue23}.

\section{\label{sec:emission_field}Emission of electric field at the dish's surface}

\begin{figure}
    \centering
    \includegraphics[scale=0.35]{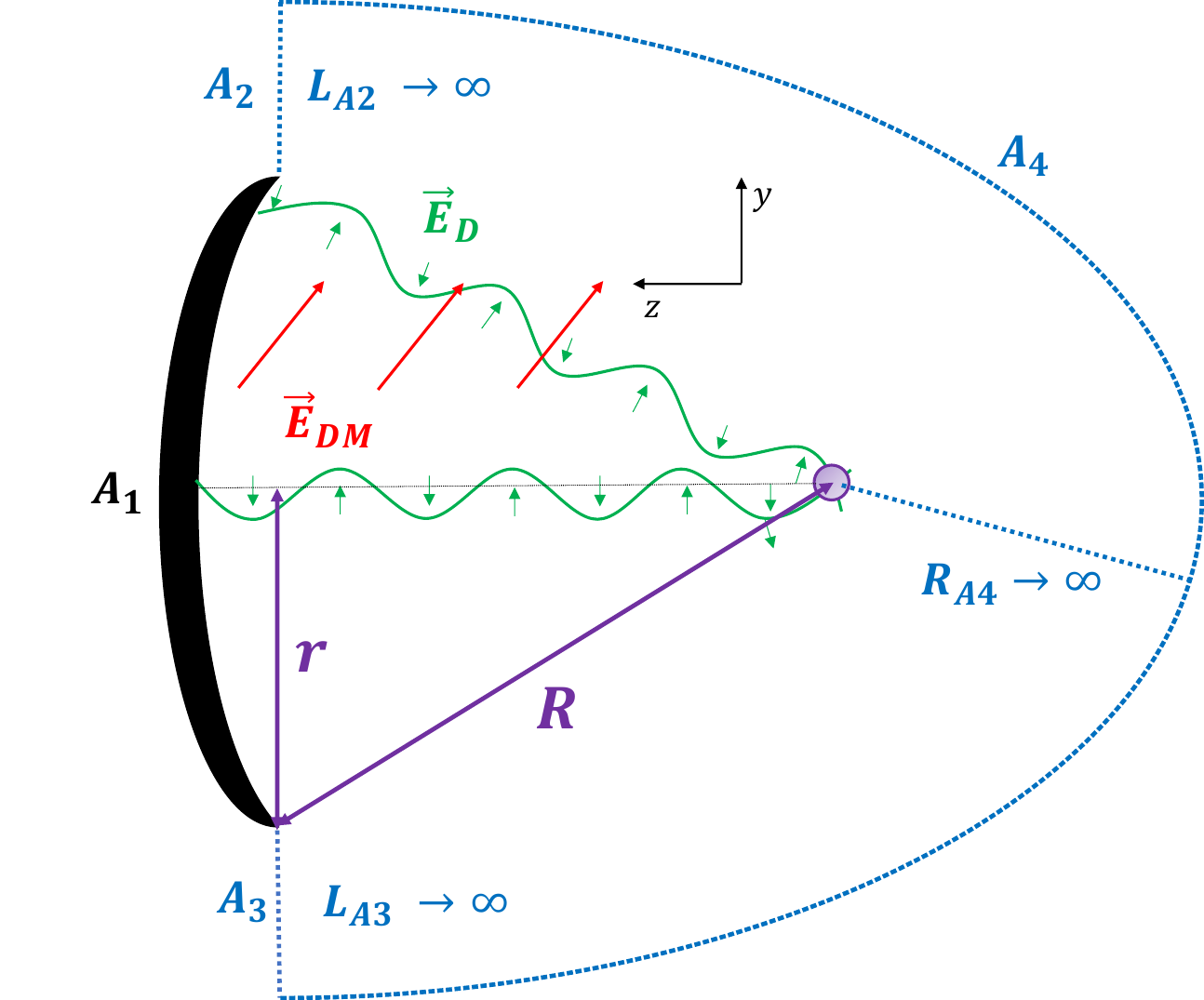}
    \caption{The oscillating standing DP electric field $\vec E_\mathrm{DM}$ is shown with red arrows. The dish emitter in black, with curvature center shown in purple, acts as reflector and emits a classic propagating electric field $\vec E_\mathrm{D}$ in green. The Kirchhoff integral boundary surfaces $A$ considered for the computation of the electric field Eq.~\eqref{E_Dirichlet} are shown in blue (in addition to the dish, surface $A_1$). It consists of the junction of a semi-sphere centered on the curvature center of the dish with infinite radius (surface $A_4$), and two additional vertical infinite plane surfaces to close the boundary surface (surfaces $A_2$ and $A_3$). This closed surface is chosen such that the contributions from surface $A_2$, $A_3$ and $A_4$ are zero at the curvature center of the dish.}
    \label{fig:SHUKET_Kirchhoff}
\end{figure}

We consider a dish of radius $r$ and curvature radius $R$, as depicted in Fig.~\ref{fig:SHUKET_Kirchhoff}.  We use a spherical coordinate system whose origin is at the curvature center of the dish such that a point on the dish is identified by $(\theta,\varphi)$. The value of the vector $\vec Y$ on the dish is denoted by $\vec Y_D(\theta,\varphi)$. The dish is surrounded by a standing electric field induced by the DP as described by Eq.~(\ref{eqs:EDM}). The dish will act as an EM reflector and as a response to the DP electric field, it will emit a standard electric field whose frequency is identical to the one from the DP field and that will propagate perpendicularly to the dish's surface. More precisely, boundary conditions $\vec E_\mathrm{tot,\parallel} = 0$ at the  dish's surface requires that the EM field emitted by the dish can be written as 
\begin{equation}
  \vec E_{D}(\theta,\varphi,t)=\Re\left[i\chi\omega \vec Y_{\parallel,D}(\theta, \varphi) e^{-i\omega t}\right] \, ,
  \label{eq:E_out}
\end{equation}
on the dish's surface. In this expression,  $\vec Y_{\parallel,D}(\theta,\varphi)$ is the projection of $\vec Y_D(\theta,\varphi)$ on the dish's surface such that the total electric field parallel to the dish vanishes at the mirror's surface. 

The time averaged total electromagnetic power generated by the dish as a reaction to the DP can be computed as \cite{Horns}
\begin{align}
  P_\mathrm{dish} &= \alpha^2 A_\mathrm{dish} \chi^2\rho_\mathrm{DM} c  \, ,
  \label{power_emit_dish}
\end{align}
where $\alpha$ is related to the polarization of the DP field and in particular $\alpha=\sqrt{2/3}$ for a DP field whose polarization is randomly distributed \cite{Horns}. In the following two sections, we wish to know what power is received by an electromagnetic antenna located at the curvature center of the dish. 

\section{\label{sec:propag_field}Propagation of the field from the dish to the antenna}
In this section, we will derive the expression of the total electric field induced by the reflection of the DP field by the dish at a given location. In Sec.~\ref{sec:Kirchhoff}, we will give a brief overview of the Kirchhoff method and shows that it cannot be directly used to get an analytical expression of the electric field for the geometry considered in Fig.~\ref{fig:SHUKET_Kirchhoff}. For this reason, we will decompose the problem into 2 sub-problems. First, we will propagate the field from the dish to the plane closing the dish (displayed in orange in Fig.~\ref{fig:Dish_plane_Vinet}) using an approximation of the Kirchhoff method valid for thin optical element as presented in Sec.~\ref{sec:dish_plane}. Subsequently, in Sec.~\ref{sec:plane2receptor}, we will propagate the electric field from the plane to the position of the detector using the Kirchhoff method.

\subsection{Kirchhoff integral}\label{sec:Kirchhoff}

The Kirchhoff's theorem allows one to compute a field quantity at a given position $\vec x$ by computing an integral over a closed surface around $\vec x$ \cite{Jackson}. 

In this formalism, the temporal dependence of the field is separated from its spatial dependence. Therefore, we decompose the emitted electric field at the dish's surface  as 
\begin{align}
\vec E_D(\vec x',t) &=  \Re[\vec U_D(\vec x')e^{-i\omega t}]\, , \label{eq:Uout}
\end{align}
where the complex function $\vec U_D(\vec x')$ denotes the spatial part of the field and $\vec x'$ is a point on the dish's surface.
We now consider a closed surface, denoted $A$, which encloses the point $\vec x$ where the value of the field is calculated.

From Kirchhoff integral's theorem, the general expression of $\vec U_D$ at this point $\vec x$ is \cite{Jackson},
\begin{align}\label{vector_Kirchhoff}
\vec{U}_D(\vec{x}) &= \int_{A} \mathrm{dS}' \left(\vec{U}_D(\vec{x'})(\hat{n}'\cdot \vec \nabla G(\vec{x},\vec{x}')) - \right. \,\nonumber\\
&\left.G(\vec{x},\vec{x}')(\hat{n}' \cdot \vec \nabla) \vec{U}_D(\vec{x'})\right)\, .
\end{align}
where $\hat n'$ is a unit vector normal to the surface $A$ directed inwardly, the derivatives are with respect to the emission coordinates $x'$ and $G$ is a Green function, appropriately defined for the situation.

In the situation depicted in the previous section, from Eq.~\eqref{eq:E_out}, we know exactly the components of the electric field at the surface of the emitting dish. Therefore, an appropriate Green function is the Dirichlet Green function $G_D$ \cite{Jackson}, defined as 
\begin{align}
G_D(\vec{x},\vec{x}')= 0 \, , \quad \forall \: \vec x' \in A \, . 
\end{align}
Then, Eq.~\eqref{vector_Kirchhoff} becomes
\begin{equation}\label{E_Dirichlet}
\vec U(\vec{x}) = \int_{A} \mathrm{dS}' \left(\vec U(\vec{x'}) (\vec \nabla G_D(\vec{x},\vec{x}')\cdot \hat{n}')\right)\, .
\end{equation}

In the following of the paper, unless otherwise specified, whenever we mention electric fields, we refer to $\vec U$, i.e its spatial part, following Eq.~\eqref{E_Dirichlet}.

In the present situation, the closed surface $A$ is appropriately defined by the junction of the dish ($A_1$), a surface with radius $R\rightarrow \infty$  ($A_4$) and two additional plane surfaces along $(x,y)$ plane to close the surface ($A_2$ and $A_3$), as described by the blue lines in Fig.~\ref{fig:SHUKET_Kirchhoff}. This  surface has the property that the only non-vanishing term in the Kirchoff integral from Eq.~(\ref{E_Dirichlet}) is the one related to the surface $A_1$, i.e.  only the field emitted by the dish will contribute.

Our system is therefore composed of a spherical dish and Eq.~\eqref{E_Dirichlet} requires the derivation of a Dirichlet Green function for a portion of sphere. However, no analytical expression has been derived (yet) in the literature for such geometry, which makes it impossible to find an exact analytical solution for our problem. Instead, we will decompose the problem into two sub-problems: using an approximation, we will propagate the electric field from the dish to the plane closing the dish (see Fig.~\ref{fig:Dish_plane_Vinet}) and then, using the Kirchhoff integral, from the plane up to the detector.

\subsection{\label{sec:dish_plane}Propagation of the electric field from the dish to the plane}

\begin{figure}
    \centering
    \includegraphics[scale=0.22]{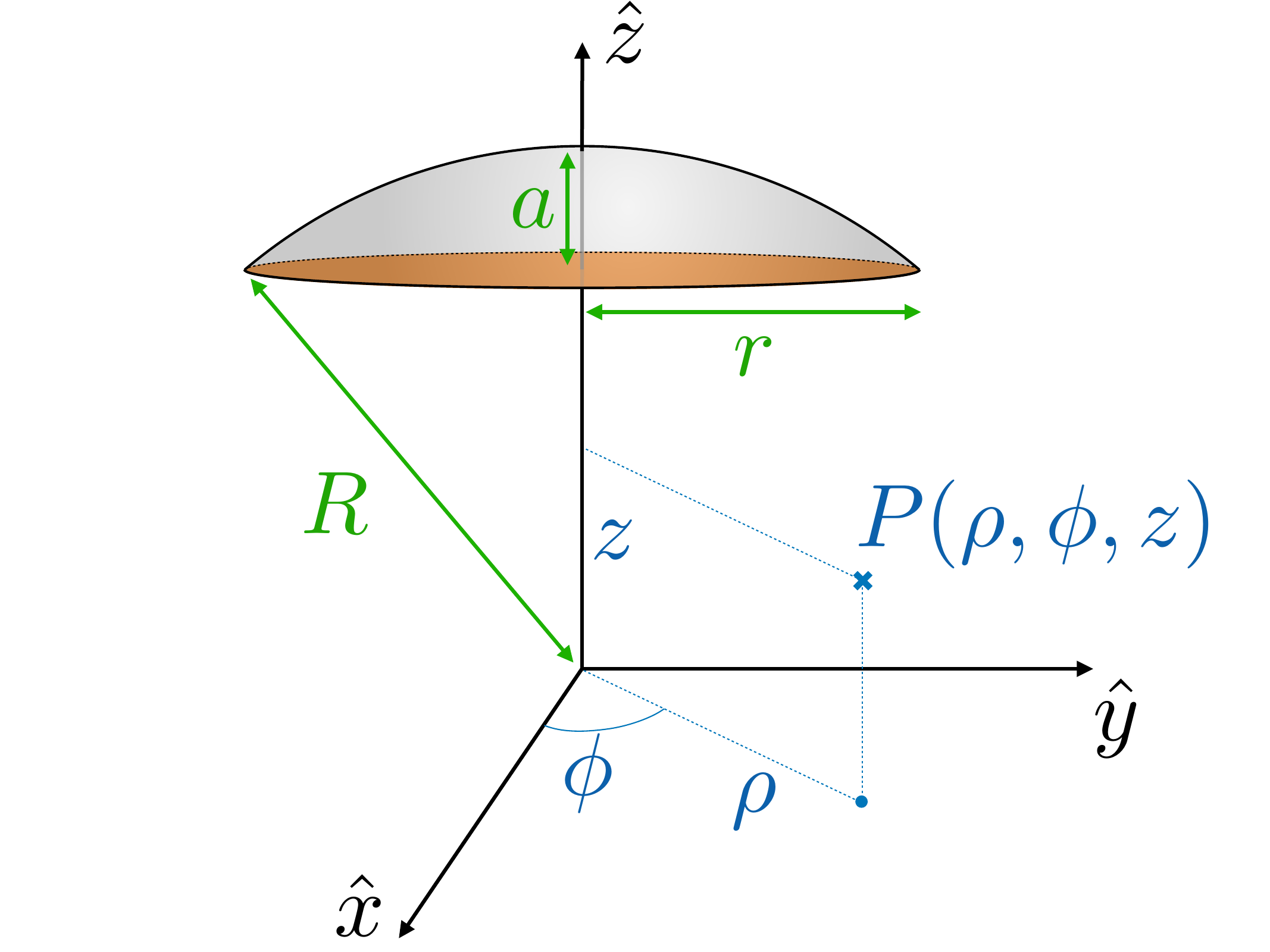}
    \caption{To compute the electric field induced by the dish at a location $P$ at cylindrical coordinates $(\rho,\phi,z)$, we first propagate the  electric field emitted by the spherical dish (in light grey) into the fictional plane (in orange) following Eq.~\eqref{Vinet}. Then, we propagate this electric field to the receiver using the Kirchhoff integral. This procedure is valid only for a spherical dish with low curvature, i.e whose radius $r$ is much smaller than its curvature radius $R$.}
    \label{fig:Dish_plane_Vinet}
\end{figure}

This situation is depicted in Fig.~\ref{fig:Dish_plane_Vinet}, where the distance between the fictional plane (orange on the figure) and the dish at coordinates $x=y=0$ is noted $a$.

As mentioned above, it is not possible to find an analytical solution to Eq.~(\ref{E_Dirichlet}) for this particular geometry. Instead, we use the \textit{thin optical element} approximation presented in details in Sec. 2.2.7 from \cite{Vinet}. This approximation is valid if the two following conditions are fulfilled:
\begin{itemize}
    \item The transverse propagation modes $p,q$ need to be much smaller than the longitudinal one $k$. In our case, the transverse modes are determined by the spatial extent of the dish antenna in the $x-y$ plane $\Delta x, \Delta y$, with $p \sim q \sim 1/\Delta x \sim 1 \: \mathrm{m}^{-1} \ll \omega/c\sim k$. In addition, the  galactic velocity distribution in the DM halo induces a transverse contribution \cite{An23}, of the order of $\omega v_\mathrm{DM}/c^2 \sim 3 \times 10^{-3} \: \mathrm{m}^{-1} \ll \omega/c\sim k$. Therefore, this condition is fulfilled.
    \item The dish needs to have a low curvature, or in other words, the radius is much smaller than the curvature radius $r \ll R$. Therefore, in the following, we will restrict ourselves to low curvature dish emitter. 
\end{itemize}
Under these conditions, the Kirchhoff integral Eq.~\eqref{E_Dirichlet} simply reduces to \cite{Vinet}
\begin{align}
\vec U_P\left(x,y,z_\mathrm{plane}\right) &= -e^{ikf(x,y)}\vec U_D\left(x,y,z_\mathrm{dish}\left(x,y\right)\right)\, ,
\label{Vinet}
\end{align}
where $\vec U_D$ is the electric field emitted by the dish and $\vec U_P$ is the electric field at the location of the closing plane. Furthermore, the function  $f(x,y)=z_\mathrm{dish}\left(x,y\right)-z_\mathrm{plane}$ is the surface equation of the dish. In other words, within the approximation of thin optical element, the curvature of the dish translates into simple phase factors determined by the distance between the closing plane and the dish \cite{Vinet}. 

We now introduce a cylindrical coordinate system ($\rho,\phi,z$), see Fig.~\ref{fig:Dish_plane_Vinet}. 
Because of the cylindrical symmetry, the surface equation that writes
\begin{align}
    f(x,y)&=f(\rho) = \sqrt{R^2-\rho^2}-R+a \approx \frac{r^2 -\rho^2}{2R} \, ,\label{eq:f}
\end{align}
depends only on the cylindrical coordinate $\rho$, where we used $a \approx r^2/2R$, since the dish is a portion of sphere, with small curvature. 

Inserting the spatial part of the field on the dish from Eq.~\eqref{eq:E_out} into Eq.~(\ref{Vinet}), we can now estimate the expression of the field in this fictional plane as
\begin{equation}\label{eq:E_plane}
\vec U_P(\rho,\phi,z_\mathrm{plane})=-i\chi \omega e^{ikf(\rho)}\vec Y_{\parallel, D}(\rho,\phi,z_\mathrm{plane}+f(\rho))\, ,
\end{equation} 
where $\rho\leq r$ and $z_\mathrm{plane}=R-a$ is the $z$-coordinate of the plane. This expression is only valid in the thin optical element approximation, whose conditions are detailed at the beginning of this section.

\subsection{\label{sec:plane2receptor}Propagation of the electric field from the plane to the receiver}

We can now focus on the propagation from an emitting plane, a subject that has been vastly treated in the literature, e.g in \cite{Jackson, Vinet}. The idea is to propagate the field $\vec U_P$ from the fictional plane $z=z_\mathrm{plane}$ to any point at $z<z_\mathrm{plane}$ using the Kirchhoff integral from Eq.~(\ref{E_Dirichlet}) by using the appropriate Dirichlet Green function that vanishes on the plane. In the case of a plane emitter, such a Dirichlet Green function is given by
\begin{align}
G_D(\vec x,\vec x')=\frac{e^{ikL'}}{4\pi L'} -\frac{e^{ikL''}}{4\pi L''}\, ,
\label{Dirichlet}
\end{align}
where $L'=\left|\vec x-\vec x'\right|$ and $L''=\left|\vec x-\vec x''\right|$ where $x''$ is the point symmetrical to $x'$ with respect to the plane \cite{Jackson,Vinet}. One can easily show that this property leads to $L'=L''$ on the plane and therefore the required condition $G_D(\vec x, \vec x_\mathrm{plane}) = 0$ is satisfied.

One can plug the expression of the Dirichlet Green function from Eq.~\eqref{Dirichlet} with $z'=z_\mathrm{plane}$ and of the electric field on the fictional plane from Eq.~(\ref{eq:E_plane}) in the Kirchhoff integral from Eq.~\eqref{E_Dirichlet} to find (the derivation is provided in Appendix.~\ref{ap:Green_derivative})
\begin{widetext}
\begin{align}
\vec U_\mathrm{dish}(\rho,\phi, z) &\approx -\frac{i\chi \omega \Delta z}{2\pi}\int_0^r  d\rho' \rho'e^{ikf(\rho')} \int_0^{2\pi}d\phi' \frac{ikL-1}{L^3}e^{ikL}\vec Y_{\parallel, D}(\rho',\phi',f(\rho')+R-a) \, ,
\label{E_field_complete}
\end{align}
\end{widetext}
with $(\rho, \phi, z)$ the cylindrical coordinate of the reception point, $f(\rho')$ provided by Eq.~(\ref{eq:f}),
\begin{subequations}
    \begin{align}
    L &= \sqrt{\rho^2+\rho'^2-2\rho\rho'\cos(\phi-\phi')+(\Delta z)^2} \label{eq:L}\, ,
\end{align}
and 
\begin{equation}
    \Delta z=z-R+a \, \\
\end{equation}
\end{subequations} the difference of $z$-coordinates between the plane emitter and the receiving point (both $z$ and $\Delta z<0$, see Fig.~\ref{fig:Dish_plane_Vinet}). This result provides the expression of the electric field induced by the DP field reflected by the dish anywhere in space under the approximation that the dish is a thin optical element. In particular, this formula includes diffraction effects that were implicitly neglected in previous studies \cite{Horns}. This integral is generally not solvable analytically and some approximations like the far field approximations (FFA), or small curvature approximations might be made to simplify it (see Section \ref{sec:SHUKET} for an example).

\section{\label{sec:detection_field}Detection of the electric field with a horn antenna}

While the previous section was devoted to the emission and propagation of the electric field induced by the dark photon and enhanced by the dish, we will now focus on the detection of this electric field.  In this paper, we consider the detection system to be a horn antenna of long side $A$ and small side $B$.
At the output of the antenna is located a resistance $R_0$ such that when an oscillating electric field is applied to the horn antenna, it is translated into a measurable voltage.

We will compare two different approaches to estimate this measurement: the first one is based on an analytical calculation which computes the overlap integral between the electric field at the location of the antenna and the antenna mode ; the second one consists in using the antenna gain factor provided by the antenna's manufacturer.

\subsection{Computation using the modes overlap}

\begin{figure}
    \centering
    \includegraphics[scale=0.22]{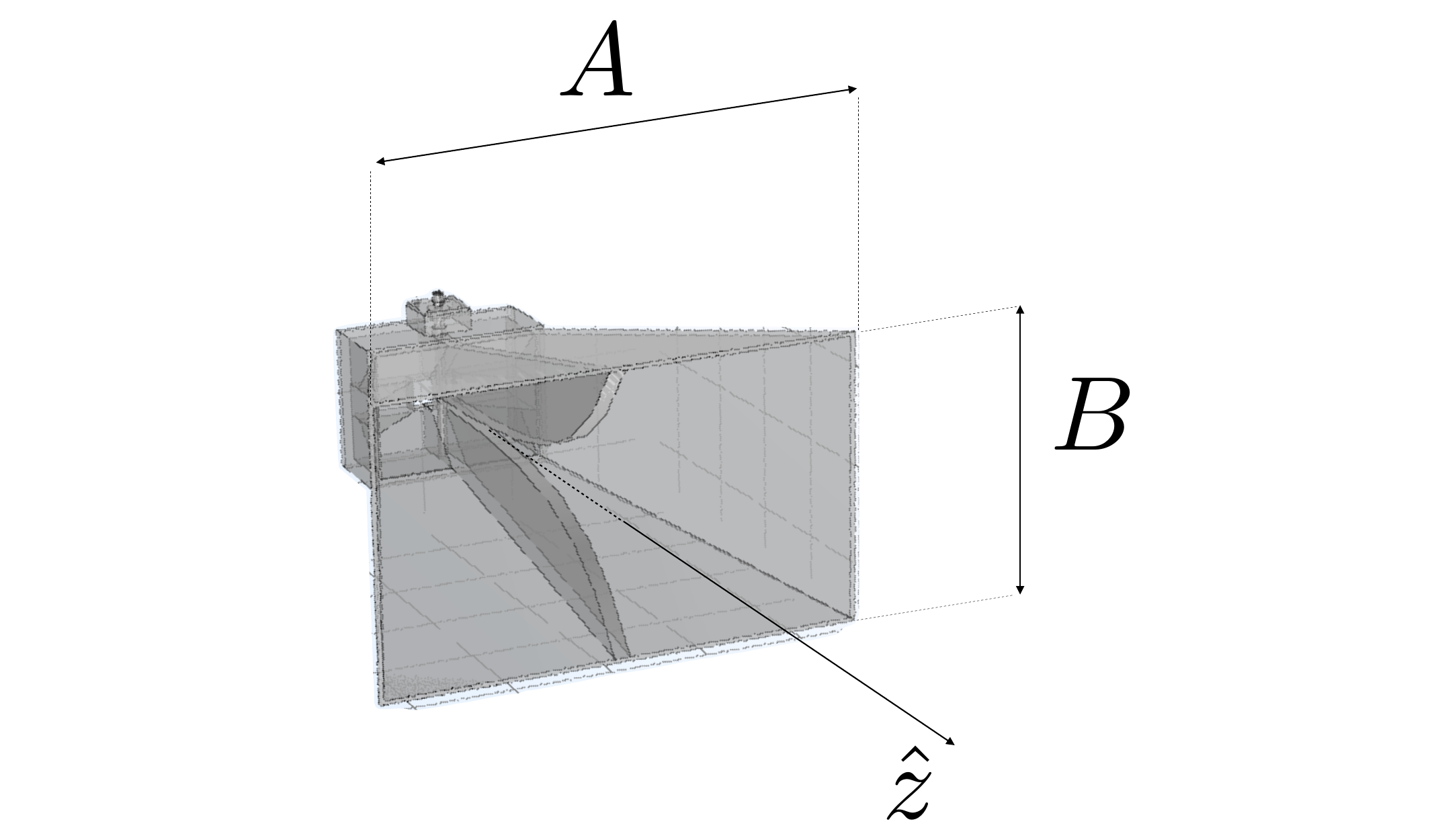}
    \caption{Rectangular horn antenna and definition of its physical surface aperture.}
    \label{fig:antenna}
\end{figure}

Let us first consider an  antenna of internal resistance $R_0$ as an emitter by applying  an oscillating voltage $V(t) = V_0 \cos (\omega t)$ to its terminals. The reciprocity theorem ensures that this will be equivalent to considering the antenna as a receiver. As a result, the antenna will emit an electric field predominantly in the TE$_{10}$ mode, i.e. polarized parallel to the small side of the rectangular horn, which is the most widely used fundamental mode for pyramidal horn antennas \cite{Balanis05}. Therefore, in a coordinate system with the $z$-axis perpendicular to the surface aperture of the antenna, as shown in Fig.~\ref{fig:antenna}.
The electric field generated by the antenna can be written as
\begin{subequations}
\begin{align}\label{eq:E_mode_ant}
\vec E_\mathrm{ant}(t,x,y,z) &= \Re\left[V_\mathrm{ant} \vec M_\mathrm{ant}(x,y) e^{i(kz-\omega t+\phi)}\right] 
\end{align}
where $\phi$ is the phase of the electric field and $V_\mathrm{ant}$ characterizes the amplitude of this field\footnote{Energy conservation implies that $V_0^2/2R_0=V_\mathrm{ant}^2/2Z_0$.}.  In general, horn antennas make an axial detection of the field, i.e. only on one single axis, the small axis of the rectangular horn, the $\hat y$ axis. The real antenna mode can be written as
\begin{align}\label{eq:mode_ant}
\vec M_\mathrm{ant}(x,y) &= m_{TE_{10}}\hat{y}\cos\left( \frac{\pi x}{A_\mathrm{eff}} \right) \,,
\end{align}
\end{subequations}
where $A_\mathrm{eff}$ is the effective width and $m_{TE_{10}}$ (dimension L$^{-1}$) is a normalisation factor that ensures the mode is normalized $\int dS_\mathrm{eff} |\vec M_\mathrm{ant}(x,y)|^2 = 1$ on the effective aperture of the antenna, i.e.
\begin{equation}\label{eq:m_value}
    m_{TE_{10}} = \sqrt{\frac{2}{S_\mathrm{eff}}} \,.
\end{equation}
The effective aperture of the antenna $S_\mathrm{eff}$ depends on the frequency $f=\omega/2\pi$ of the field and in general differs from the physical aperture of the antenna. Indeed, the effective area of the horn, which can be expressed as the product of an effective width $A_\mathrm{eff}$ with an effective height $B_\mathrm{eff}$, depends on the frequency as \cite{Balanis05}
\begin{align}
S_\mathrm{eff}(\omega) &= A_\mathrm{eff}(\omega)B_\mathrm{eff}(\omega)=\frac{e_r\pi G(\omega)c^2}{\omega^2} \, ,
\label{eq:effective_surface}
\end{align}
where $e_r$ is the realistic efficiency of the antenna, which represents all the losses inside the antenna, including e.g. reflections inside the antenna, and where $G(\omega)$ is the frequency dependent gain of the antenna \cite{Balanis05}. The time-averaged electromagnetic power generated by the antenna is given by
\begin{equation}\label{eq:Pout}
    P_\mathrm{out} = \frac{1}{2Z_0} \int dS_\mathrm{eff} \left|\vec E_\mathrm{ant}\right|^2 =\frac{V_\mathrm{ant}^2}{2Z_0} \, ,
\end{equation} 
where $Z_0=376.7 \: \Omega$ is the impedance of vacuum.

We will now invert this reasoning and consider the antenna as a receiver that will output a voltage in response to an electric field $\vec E(t,x,y,z)=\Re[\vec U(x,y,z)e^{-i\omega t}]= \left|\vec U\right|\hat e_U\cos(\omega t +\varphi)$, where $\varphi$ is an irrelevant phase, $\hat e_U$ is the unit polarization vector of $\vec U$, $\left|\vec U\right|=\left(\vec U\cdot \vec U^\dagger \right)^{1/2}$ denotes the complex modulus. The antenna will  measure electric fields that are propagating into the $\hat z$ direction and matching its mode $\vec M_\mathrm{ant}(x,y)$. More precisely, by multiplying Eq.~(\ref{eq:E_mode_ant}) with $\vec M_\mathrm{ant}$, one show that the antenna will output a voltage proportional to
\begin{align}\label{eq:antenna_output_gen}
    &V_\mathrm{ant}(z)\cos(\omega t+\phi'(z))=\cos(\omega t+\varphi) \, \\
    &\int dS_\mathrm{eff}(\omega) \left|\vec U(x,y,z)\right| \hat e_U \cdot \vec M_\mathrm{ant}(x,y) \, \nonumber ,
\end{align} 
where $\phi'(z)=-kz-\phi$ and the 2D integral is carried out in the $x-y$ plane at the $z$-coordinate of the antenna.

Let us factorize the electric field generated by the dish at the location of the antenna provided by Eq.~(\ref{E_field_complete}) in a voltage factor $V_\mathrm{dish}$, a real mode factor $\vec M_\mathrm{dish}$ and a plane wave factor, such that it takes the following form in the antenna aperture plane $(x,y)$
\begin{eqnarray}\label{eq:E_mode_dish}
\vec {E}_\mathrm{dish}(t,x,y,z) &= V_\mathrm{dish} \vec{M}_\mathrm{dish}(x,y,z) \cos(\omega t+\phi'(z)) \,\nonumber \\
&\equiv \left|\vec U_\mathrm{dish}(x,y,z)\right|\hat e_U\cos(\omega t+\varphi) \, ,
\end{eqnarray}
where $\vec{M}_\mathrm{dish}(x,y,z)$ is normalized over the infinite surface of the antenna plane $S_\infty$ at $z=z_\mathrm{ant}$ i.e
\begin{align}
    \int dS_\infty |\vec{M}_\mathrm{dish}(x,y,z_\mathrm{ant})|^2 = 1 \, .
    \label{mode_dish_norm}
\end{align}
The constant amplitude $V_\mathrm{dish}$ in Eq.~\eqref{eq:E_mode_dish} can be computed through energy conservation: the total energy generated by the dish should equal the energy received on the infinite antenna plane $S_\infty $. Then, using Eqs.~\eqref{power_emit_dish} and \eqref{mode_dish_norm}, we have
\begin{subequations}
\begin{align}
 P_\mathrm{dish} &= \int dS_\infty \frac{\epsilon_0\left|\vec U_\mathrm{dish}\right|^2 c}{2} =\frac{V_\mathrm{dish}^2}{2Z_0} \,\\
\Rightarrow V_\mathrm{dish} &= \sqrt{\frac{4A_\mathrm{dish}\rho_\mathrm{DM}}{3\epsilon_0}}\chi \label{eq:Vdish} \, ,
\end{align}
\end{subequations}
where $\epsilon_0$ is the permittivity of vacuum.

Then, in our experimental scheme, the time averaged power generated by the horn antenna is obtained by combining Eqs.~\eqref{eq:Pout}, ~(\ref{eq:antenna_output_gen}) and (\ref{eq:E_mode_dish}) and is provided by
\begin{equation}\label{eq:overlap_int}
P_\mathrm{rec}(z_\mathrm{ant},\omega) = \frac{V_\mathrm{dish}^2}{2Z_0}\left(\int dS_\mathrm{eff}(\omega) \vec{M}_\mathrm{dish}\cdot \vec{M}_\mathrm{ant}\,\right)^2 \, ,
\end{equation}
where $\vec M_\mathrm{dish}$ depends explicitly on $z_\mathrm{ant}$ and where the integral is performed at every position $(x,y)$ on the effective surface of the horn antenna. 
Then, the ratio between the power measured by the horn antenna and the total power emitted by the dish is simply given by
\begin{align}
\gamma(z_\mathrm{ant},\omega)_\mathrm{Overlap} &= \left(\int dS_\mathrm{eff}(\omega) \vec{M}_\mathrm{dish}\cdot \vec{M}_\mathrm{ant}\right)^2 \, ,
    \label{ratio_powers}
\end{align}
where $\vec{M}_\mathrm{dish}$ directly depends on the Kirchhoff integral Eq.~\eqref{E_field_complete}.

\subsection{Computation using the antenna factor\label{sec:antenna_factor_gen}}

Another approach to estimate the output of the horn antenna is to consider the antenna factor (AF), a characteristic of the antenna provided by the manufacturer. 

The AF is defined by considering an incident plane wave, or in other words, an incoming electromagnetic field whose mode is constant over the aperture of the antenna. It is defined by (see e.g. \cite{Mclean02})
\begin{align}
\mathrm{AF}(\omega) &= \frac{\left|\vec U_\mathrm{dish}(\omega)\right|}{V_0} = \frac{V_\mathrm{dish} M_\mathrm{dish}}{V_0}  \, , 
\end{align}
where $V_\mathrm{0}$ is the generated tension, $\vec U_\mathrm{dish}$ the amplitude of the incoming electric field, $\vec M_\mathrm{dish} = M_\mathrm{dish} \hat e_U$ from Eq.\eqref{eq:E_mode_dish}, with ${M}_\mathrm{dish}$ the value of the constant mode of the electric field on the effective antenna area. The AF depends on the frequency of the incident field. 
It is measured experimentally and therefore takes into account any loss inside the antenna, represented by the $e_r$ parameter. 

Using energy conservation, we  immediately get the expression of time-averaged generated power of the antenna
\begin{align}
P_\mathrm{rec}(z_\mathrm{ant},\omega) = \frac{V_0^2}{2R_0}=\frac{V^2_\mathrm{dish}M^2_\mathrm{dish}(z_\mathrm{ant},\omega)}{2 R_0 \mathrm{AF}(\omega)^2} \, .
\end{align}
However, the definition of the antenna factor assumes perfectly aligned polarization modes \cite{Mclean02}, which is not necessarily true in our case. Indeed, we assume a antenna polarization mode along the $\hat y$ axis while the electric field emitted by the dish has its polarization lying in the $(x,y)$ plane. 
 If we assume the DP polarization to be isotropically distributed, the polarization of the electric field emitted by the dish has a cylindrical symmetry and a linearly polarized antenna will be sensitive only to half of the power from the electric field.
 Taking into account this additional factor, the ratio $\gamma$ becomes
\begin{align}
\gamma(z_\mathrm{ant},\omega)_\mathrm{AF} &= \frac{Z_0 M^2_\mathrm{dish}(z_\mathrm{ant},\omega)}{2 R_0 \mathrm{AF}(\omega)^2} \, .
\label{ratio_AF}
\end{align}

\section{\label{sec:SHUKET}A practical example: the case of the  SHUKET experiment}

We now focus on a practical example to illustrate how the Kirchhoff integral theorem and the overlap of modes affect the expected power received by an antenna from a spherical dish emitter.

We consider the setup of the SHUKET experiment \cite{SHUKET}, which has been used to search for a DP in the frequency range $5-6.8$ GHz using a spherical dish and a horn antenna for the detection. In the analysis of this experiment, following \cite{Horns}, it has been assumed that all the power emitted by the dish is received by the antenna. Following the method derived in the previous section, we estimate the power received by the antenna and estimate the power lost compared to this ideal situation, for the mean DP frequency, i.e $f=6$ GHz.

\subsection{Propagation of the electric field from the dish through the antenna}\label{sec:shuket_prop}

The spherical dish used in \cite{SHUKET} has a curvature radius of $R=32$ m and an area of $A_\mathrm{dish} = 1.2$ m$^2$, implying a radius of $r \approx 0.618$ m (see Fig.~\ref{fig:SHUKET_Kirchhoff}). Then, $R \gg r$ (or equivalently the dish-fictional plane distance $a\sim 6 \times 10^{-3}$ m  in the case of SHUKET is way smaller than $r$, i.e. $a\ll r$) which ensures that the low curvature approximation is valid. 

The expression of  the  electric field emitted by the dish at its surface is given by Eq.~(\ref{eq:E_out}) where $ \vec Y_{\parallel,D}$ has the form 
\begin{align}
 \vec Y_{\parallel,D} = \vec Y - \left(\vec Y\cdot \hat e_r\right)\hat e_r \approx \begin{pmatrix}
      Y_x \,\\
      Y_y \,\\
      0
  \end{pmatrix} + \mathcal{O}(\theta)\, \label{field_polarization}, 
\end{align}
with $\theta, \phi$ corresponding to a spherical coordinate system centered in dish's curvature center and where we have used the notation $\vec Y=(Y_x,Y_y,Y_z)$ in the cartesian coordinate system depicted in Fig.~\ref{fig:SHUKET_Kirchhoff} and where we used the low curvature approximation at the last step\footnote{The relative error induced by the neglect of $\mathcal{O}(\theta)$ is $\sim$ 1\%.}. This means that at leading order in $\theta$, the polarization of the emitted electric field does not depend on the dish coordinate, and is only polarized in the $x-y$ plane.

For this configuration, the power emitted by the dish is independent of the mass of the DP and is given by Eq.~\eqref{power_emit_dish}\footnote{Note there is a factor 2 discrepancy with what \cite{SHUKET} indicated.} 
\begin{align}\label{power_SHUKET}
    P^\mathrm{SHUKET}_\mathrm{dish} &= 1.73 \times 10^{-20} \left(\frac{\chi}{10^{-12}}\right)^2 \mathrm{ \ W} \, ,
\end{align}
where we considered a DM local energy density $\rho_\mathrm{DM} = 0.45$ GeV/cm$^3$. 

The calculation presented in Sec.~\ref{sec:propag_field} is based on the thin optical element approximation, which is valid only for low curvature dish $R\gg r$, see the discussion in Sec.~\ref{sec:dish_plane}. In Appendix~\ref{test_Vinet}, we show that this approximation is valid in the SHUKET setup and we quantify numerically the relative error on the electric field to be of order $10^{-4}$ of such approximation. Additionally, the calculation in Appendix.~\ref{test_Vinet} shows that the small curvature approximation and Eq.~\eqref{field_polarization} are valid in our regime.

The electric field induced by the dish at the location of the antenna is provided by Eq.~(\ref{E_field_complete}), which is not solvable analytically. In order to simplify it and get an analytical expression, we will make different  approximations:
\begin{itemize}
    \item  The distance between the fictional plane and the antenna needs to be much larger than the typical size of the dish, i.e. $\left|\Delta z\right| \gg \rho' \leq r$. 

    \item  The distance between the fictional plane and the antenna needs to be much larger than the typical size of the antenna, i.e. $\left|\Delta z\right| \gg \rho$.

    \item  The last approximation is known as the far field approximation (FFA). For the DM Compton frequency under consideration, i.e $f=6$ GHz, $kL \sim k |\Delta z| \gg 1$ and we can safely neglect the factor -1 in Eq.~(\ref{E_field_complete}).
\end{itemize} 

The first two approximations simplify the distance $L$ from Eq.~\eqref{eq:L} between any point on the fictional plane $(\rho',\phi',R-a)$ and any point on the antenna (whose center is located at the origin of our coordinate system) $(\rho,\phi,0)$ to
\begin{align}\label{eq:L_low_curv}
L &\approx |\Delta z| + \frac{\rho^2+\rho'^2-2\rho\rho'\cos(\phi-\phi')}{2|\Delta z|} \, .
\end{align}

Using this expression of $L$ as well as the FFA, we can express Eq.~\eqref{E_field_complete} as 
\begin{subequations}
\begin{align}
   \vec U^\mathrm{SHUKET}_\mathrm{dish}(\rho, \Delta z) &\approx \frac{\omega^2\chi e^{i\Phi(\rho,\Delta z)}}{\Delta zc}\begin{pmatrix}
Y_x\\
Y_y\\
0
\end{pmatrix}\, \\
&\times \int_0^r d\rho' \rho' e^{-i\varphi(\rho',\Delta z)}J_0\left(\frac{k\rho\rho'}{|\Delta z|}\right) \, \nonumber ,\label{E_field_FFA} 
\end{align}
where $J_0$ is the Bessel function of the first kind of order 0 and where the integral is performed over the radius of the fictional plane which closes the dish, where the dependence on the angle $\phi$ disappeared by spherical symmetry and with
\begin{align}
    \varphi(\rho',\Delta z)&= \frac{k\rho'^2}{2}\left(\frac{1}{R}+\frac{1}{\Delta z}\right)\,\\
    \Phi(\rho,\Delta z) &= k\left(\frac{r^2}{2R} - \Delta z - \frac{\rho^2}{2\Delta z}\right) \, .
\end{align}
\end{subequations}
One can find an analytical solution for the last integral in the case where $z_\mathrm{ant}=0$ (see Appendix.~\ref{FFA_analytic})
\begin{equation}
    \vec U^\mathrm{SHUKET}_\mathrm{dish}(\rho)\approx \frac{r \omega\chi}{\rho}\begin{pmatrix}
Y_x\\
Y_y\\
0
\end{pmatrix}J_1\left(\frac{k\rho r}{R-a}\right)e^{i\Phi(\rho,a-R)} \, ,
\label{E_field_final}
\end{equation}
with $J_1$ the Bessel function of the first kind of order 1. 
 
At the center of the curvature radius of the dish is located a polarized horn-antenna of physical surface $S_\mathrm{phys}=0.25 \times 0.142 \ \mathrm{m}^2$.
Numerical integration of the power from the electric field from Eq. (\ref{E_field_final}) over the physical antenna surface leads to
\begin{align}
    P_\mathrm{int}&=\int dS_\mathrm{phys}\frac{\epsilon_0 |U^\mathrm{SHUKET}_\mathrm{dish}(\sqrt{x^2+y^2})|^2 c}{2}\,\nonumber \label{S_rec_geom}\\
&\approx 2.85 \times 10^{-22} \Big(\frac{\chi}{10^{-12}}\Big)^2 \mathrm{ \ W} \, ,
\end{align}
where we assumed an emission from a plane surface in the random polarization scenario, as in Eq.~\eqref{power_emit_dish}. The ratio of the power emitted by the dish that crosses physically the antenna and the total power emitted by the dish in  the SHUKET experiment is
\begin{equation}
R = \frac{P_\mathrm{int}}{P^\mathrm{SHUKET}_\mathrm{dish}} \approx\frac{2.85 \times 10^{-22}}{1.73 \times 10^{-20}} \approx 1.6\: \mathrm{\%} \, .
\end{equation}
Note that, in the geometrical optics approximation, i.e where diffraction effects are neglected, one would obtain $R=100\: \mathrm{\%}$.
It is interesting to note that the ratio of the physical antenna's surface to the dish's surface is $\sim 3\%$, meaning that there is actually no focus of the field generated by the dish on the antenna.

This result means that, considering simply the propagation of the field from the dish to the antenna using Kirchhoff integral, the majority of the electromagnetic power is lost through diffraction, and the antenna is only able to detect a small amount of energy emitted by the dish.  

One can note that the usual criteria for diffraction effects to be negligible $d_\mathrm{dish} = 2r \gg \lambda$ is not fulfilled in this system, as the proportionality factor between the two parameters is $\sim$ 25, which explains this lack of focus.

As a cross-check of our calculations, integrating Eq.~\eqref{S_rec_geom} over the infinite antenna plane gives
\begin{align}
P_\mathrm{int, full} &\approx 1.73 \times 10^{-20} \Big(\frac{\chi}{10^{-12}}\Big)^2 \mathrm{ \ W} ,
\end{align}
which corresponds to the full power emitted $P^\mathrm{SHUKET}_\mathrm{dish}$ from Eq.~\eqref{power_SHUKET}. This validates our whole set of approximations, made throughout the derivation.

\subsection{Detection of the field using a horn antenna}\label{sec:shuket_detection}

In the previous section, we showed that most of the power emitted by the dish is already lost through propagation of the field from the dish to the antenna. As described in Sec.~\ref{sec:detection_field}, there is still a second step to consider before predicting the exact amount of energy generated by the antenna: overlap integral between incident and antenna modes. In the SHUKET experiment, this detection is made by the \textit{Schwarzbeck BBHA-9120-D} antenna, which is sensitive to electric field frequencies from 0.8 to 18 GHz. The datasheet \cite{datasheet_ant} contains several pieces of information that will be necessary for our calculations on the expected amount of power received by the antenna wires, namely the gain and the antenna factor of the antenna as function of those various field frequencies. These quantities have been measured experimentally by the manufacturer.

\subsubsection{Computation using the modes overlap}
As mentioned in Sec.~\ref{sec:detection_field},  to predict the  energy generated by the horn antenna, i.e the overlap of modes, we need to consider the effective surface of the antenna $S_\mathrm{eff}$ at the frequency we are interested in ($f$ = 6 GHz). Furthermore, we also need to find an analytical expression for the mode of the field emitted by the dish $\vec M_\mathrm{dish}(x,y)$, such that it is possible to perform the mode overlap integral Eq.~\eqref{eq:overlap_int}.

The antenna consists in a rectangular surface of area $S_\mathrm{phys} = 0.25 \times 0.142$ m$^2$,  with $A=0.25$ m the long length and $B=0.142$ m the small length. This implies that for the fundamental mode of the antenna at frequency $\omega_c$, the ratio between the long and small sides of the antenna is $R_\mathrm{AB}(\omega_c)=A_\mathrm{eff}(\omega_c)/B_\mathrm{eff}(\omega_c)=0.25/0.142\approx 1.76$.  We will make the hypothesis that the ratio $R_{AB}$ is independent of the frequency,  i.e $R_\mathrm{AB}(\omega_c) \equiv R_\mathrm{AB} = A_\mathrm{eff}(\omega)/B_\mathrm{eff}(\omega)$, for any frequency $\omega>\omega_c$, such that the effective surface of the antenna at frequency $\omega$ is simply $S_\mathrm{eff}(\omega)= A_\mathrm{eff}(\omega)B_\mathrm{eff}(\omega)=A_\mathrm{eff}^2(\omega)/R_\mathrm{AB}$. Then,  using Eq.~\eqref{eq:effective_surface}, the effective long and small lengths of the antenna at the frequency $\omega$ are simply given by
\begin{subequations}
\begin{align}
\frac{A_\mathrm{eff}^2(\omega)}{R_\mathrm{AB}} &= \frac{e_r\pi G(\omega)c^2}{\omega^2}\,\\
B_\mathrm{eff}(\omega)&=\frac{A(\omega)}{R_\mathrm{AB}} \, .
\end{align}
\end{subequations}
These two dimensions defined the effective surface area over which we need to integrate Eq.~(\ref{eq:overlap_int}) to estimate the output of the antenna.

To be able to compute the integral overlap of modes, we need an analytic expression of the field at coordinate $(\rho,z)$, which is provided by Eq.~\eqref{E_field_final}.
Then, from Eqs.~\eqref{eq:E_mode_dish} and \eqref{E_field_final}, we can separate the mode of the dish $\vec{M}_\mathrm{dish}$ from the constant amplitude $V_\mathrm{dish}$ expressed in Eq.~\eqref{eq:Vdish}, such that the mode of the dish at coordinates $(x,y,0)$ is
\begin{align}\label{dish_mode_SHUKET}
\vec{M}_\mathrm{dish}(x,y) &= \sqrt{\frac{3}{2\pi}} \frac{J_1\left(\frac{rk\rho}{R-a}\right)}{\rho\left|\vec Y\right|}\begin{pmatrix}
    Y_x\\
Y_y\\
0
\end{pmatrix} \, ,
\end{align}
with $\rho=\sqrt{x^2+y^2}$. Then, using Eqs.~\eqref{eq:mode_ant}, \eqref{ratio_powers} and \eqref{dish_mode_SHUKET} and assuming the polarization of the DP to be randomly distributed, the ratio of receiving to emitted powers is simply
\begin{align}
\gamma_\mathrm{Overlap} &= \left(\int dS_\mathrm{eff} \vec M_\mathrm{ant}\cdot  \vec M_\mathrm{dish}\right)^2 \approx 5.8 \times 10^{-4} \, ,
\label{ratio_overlap}
\end{align}
with $m_{TE_{10}} \approx 25.6$ m$^{-1}$ has been estimated from Eqs.~(\ref{eq:m_value}) and (\ref{eq:effective_surface})  using the antenna gain $G(f)=11.86$ dBi at $f=6$ GHz provided in the antenna datasheet and which has been measured experimentally by the manufacturer. In addition, we assumed $e_r=1$, i.e no loss inside the antenna, because we do not have any numerical value for it.
For this numerical value of ratio, we assumed an axial detection, with $Y_x = Y_y$ in Eq.~\eqref{dish_mode_SHUKET}.
This result means that only 0.06\% of the emitted power is actually transmitted to the antenna wires, and therefore detectable.

\subsubsection{Computation using the antenna factor}

As explained in Sec.~\ref{sec:antenna_factor_gen}, if the  electric field emitted by the dish is seen as a plane wave by the antenna or equivalently if the mode of this electric field is approximately constant over the effective surface of the antenna, we can use another  method to derive output of the experiment. 

From Eq.~\eqref{dish_mode_SHUKET}, it can be shown that the dish polarization mode is approximately constant over the effective long length of the antenna ($\sim 7$ cm) (with a deviation of $\sim 0.3\%$), with a value of $M^y_\mathrm{dish}(x,y=0) \approx 0.48 \: \mathrm{m}^{-1}$.

For a Horn antenna with internal  $R_0$=50$\ \Omega$ resistance (which is typically the case for the \textit{Schwarzbeck BBHA-9120-D} antenna), the antenna factor at frequency $\omega$ is given by \cite{Mclean02}
\begin{align}
\mathrm{AF}(f) &= \frac{9.73 \ f}{c \sqrt{G(f)}} = 49.7 \mathrm{\ m}^{-1} \, ,
\label{AF_gain}
\end{align}
which is consistent with the value given in the antenna datasheet \cite{datasheet_ant}\footnote{Since AF(dB)=$20\log_\mathrm{10}$(AF) \cite{Balanis05}.}
Then, using Eq.~\eqref{ratio_AF}, one can compute the ratio of received to emitted powers as 
\begin{align}
\gamma_\mathrm{AF} &\approx 3.6 \times 10^{-4} \, .
\label{ratio_AF_SHUKET}
\end{align}
Eqs.~\eqref{ratio_overlap} and \eqref{ratio_AF_SHUKET} disagrees with an approximate factor 1.5 difference. This difference between the two methods is most likely coming from the loss factor $e_r$ that we did not consider in the overlap integral method. Indeed, we can artificially consider $e_r \sim 0.62$ such that both methods coincide.
Therefore, the second result, obtained using the antenna factor, Eq.~\eqref{ratio_AF} is probably more realistic, as the different parameters have been experimentally measured. 

\subsection{Updated results for the SHUKET experiment}
Combining the results from the previous two subsections, we can now reevaluate the constraints on $\chi$ obtained in the SHUKET experiment \cite{SHUKET}.

\begin{figure*}
    \centering
    \includegraphics[width=\textwidth]{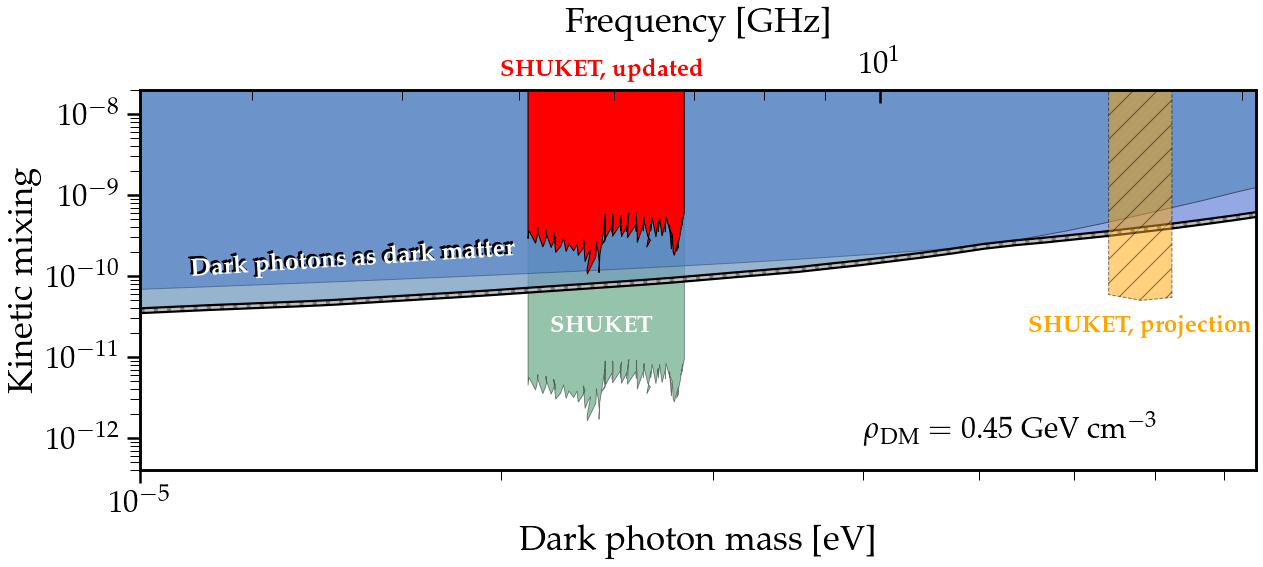}
    \caption{In green is the original constraint on the kinetic mixing parameter from the SHUKET experiment obtained in \cite{SHUKET}. In red is shown the updated constraint using the same experimental parameters and data but considering a realistic modeling of the experiment in the analysis, following Eq.~\eqref{ratio_AF_SHUKET}.  The sensitivity curve with optimized experimental parameters is shown in dashed orange, following Eq.~\eqref{ratio_opti} (from \cite{AxionLimits}).}
    \label{fig:SHUKET_sens_maj}
\end{figure*}

In Fig.~\ref{fig:SHUKET_sens_maj}, we show how both effects (diffraction and mode overlap) affect the sensitivity of the SHUKET experiment. In green is the original sensitivity curve presented in \cite{SHUKET}. Since the power received by the antenna wires is quadratic in $\chi$ coupling, the ratio Eq.~\eqref{ratio_AF_SHUKET} leads to a loss in $\chi$ of an approximate factor 53. Making the assumption that this loss factor is roughly the same over all DM frequencies to which SHUKET is sensitive, this leads to an updated sensitivity curve, shown in black in Fig.~\ref{fig:SHUKET_sens_maj}.

\subsection{Optimization of the experimental parameters to enhance the output signal}

Following the full derivation described above, it is possible to find some optimized experimental configuration such that the power received by the antenna is maximized. The experimental parameters that can easily be modified are the emitter-detector distance $L$ and the optimized DM frequency $f$ to search for with this setup\footnote{This requires an antenna appropriate for this frequency.}.

\subsubsection{Optimal frequency and distance}
The frequency of the background DM field has several impacts on the power received by the antenna. First, following Eq.~\eqref{E_field_complete}, the electric field amplitude received at coordinates $(\rho,\Delta z)$ (obtained using the Kirchhoff integral) depends highly on the frequency of the field. It can be shown numerically from Eq.~\eqref{E_field_FFA} that the focusing of the electric field improves with  the DM frequency, which is  a consequence of diffraction effects that become non negligible for low frequency, i.e. when $f d_\mathrm{dish}/c = 2fr/c \: \slashed{\gg} \: 1$. In particular, as already discussed in Sec.~\ref{sec:SHUKET}, for the size of dish and the frequency bandwidth of the horn antenna used in SHUKET, diffraction effects are always non negligible (i.e.  $f d_\mathrm{dish}/c \: \slashed{\gg} \: 1$). As a consequence, the optimal location for the horn antenna is not the center of curvature of the dish and this location becomes frequency dependent. In addition, the antenna factor (or equivalently the antenna gain) is also highly dependent on the frequency of the measured electric field. As it was shown in the previous section, the overlap integral of polarization modes contribute almost equally to the total power loss than the one from the propagation of the field. Therefore, the optimal distance $L$ between the dish emitter and the horn antenna detector where the maximum field power is transmitted is non trivial and depends on the frequency.

The goal of this section is to explore the parameter space to find the optimal frequency $f$ and distance $L$ such that the efficiency coefficient $\gamma_\mathrm{AF}$ from Eq.~\eqref{ratio_AF} is maximized. To do so, we use Eq.~\eqref{E_field_FFA} with unknown parameter $\Delta z$\footnote{Note that in order to use this equation, we must restrict ourselves to dish-horn distance $\left|\Delta z \right|\gg r$.} and Eq.~\eqref{eq:Vdish} to find the mode of the field emitted by the dish. 
We first consider only the value of the mode at $\rho=0$, and then we show that for the optimized parameters, the mode is indeed constant over the effective size of the antenna, such that the method can be used. Additionally, we interpolated the antenna datasheet \cite{datasheet_ant} to infer the value of the antenna factor as function of the frequency $\mathrm{AF}(f)$. Then, Eq.~(\ref{ratio_AF}) is used to estimate the efficiency coefficient  
\begin{align}\label{eq:gamma_AF_opt}
    \gamma(f, \Delta z)_\mathrm{AF} &= \frac{Z_0 M^2_\mathrm{dish}(\rho=0,\Delta z,f)}{2R_0\mathrm{AF}(f)^2} \, ,
\end{align}
which is estimated numerically. 

\begin{figure}
    \centering
    \includegraphics[scale=0.58]{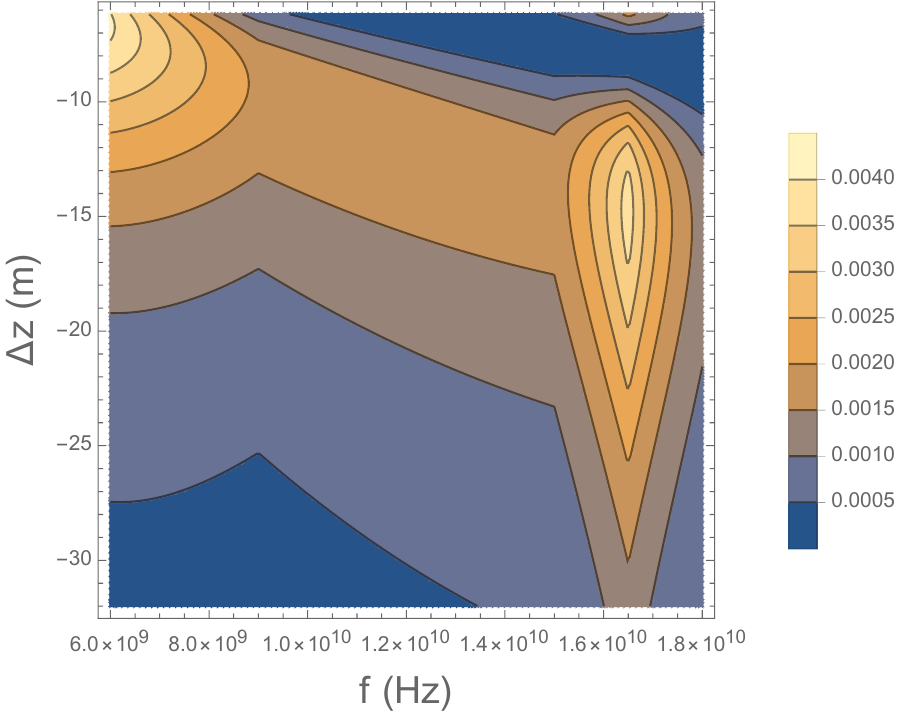}
    \caption{Efficiency coefficient $\gamma(f,\Delta z)_\mathrm{AF}$ as function of the frequency $f$ and $\Delta z$, in the frequency range $f \in [6, 18]$ GHz, of the horn antenna \cite{datasheet_ant} and for distances $|\Delta z| \gg r$. The efficiency coefficient increases for low frequencies and short distances and presents a local maximum around $(f \sim 1.65 \times 10^{10} \: \mathrm{Hz}, \Delta z \sim -15 \: \mathrm{m})$.}
    \label{fig:3D_plot_gamma}
\end{figure}

The behavior of this efficiency coefficient as a function of the DP frequency and of the distance between the dish and the receiver is shown in  Fig.~\ref{fig:3D_plot_gamma}. One can notice the increase of $\gamma$ for small frequencies and short distances, which is mainly driven by the behavior of the antenna factor. Indeed, even though the loss through diffraction effects is larger at lower frequencies, the $\gamma$ parameter also takes into account the mode matching of the antenna, which is larger at those frequencies. 

We choose a frequency band of the same size as the initial constraint from SHUKET, i.e 1.8 GHz. In addition, we restrict ourselves to frequencies larger than 10 GHz, because there is still large unexplored of the parameter space in this region, see e.g \cite{AxionLimits, Caputo}. From Fig.~\eqref{fig:3D_plot_gamma}, one can see that the optimized region for large efficiency coefficient is in the approximate range $f \in [15.5, 17.3]$ GHz. For the set of optimized frequencies $f_\mathrm{set} =\{15.5, 16.5, 17.3\}$, 
we compute the corresponding optimized distance $\Delta z_O$. For each of these configurations, we take the exact value of the antenna factor from the datasheet \cite{datasheet_ant} and we check that the mode of the dish is constant over the effective aperture of the antenna (with a deviation from a straight line of the order of $1\%$.


\subsubsection{Sensitivity analysis in the case of optimized experimental parameters}
The expected output signal from  the SHUKET experiment using the optimized experimental parameters derived in the last section can be computed using the same procedure and approximations as the ones presented in Sec.~\ref{sec:shuket_prop} and \ref{sec:shuket_detection} at the exception of one approximation that is no  longer valid. Indeed, Eq.~(\ref{E_field_final}) is obtained by assuming that $z\ll R$ which is no longer the case for the optimized distances, as it is shown in Fig.~\ref{fig:3D_plot_gamma}.  

In Appendix~\ref{FFA_analytic_opti}, we compute the power measured by the antenna in these optimized cases, starting from Eq.~(\ref{E_field_FFA}). For example, assuming the frequency $f_O=16.5$ GHz, and the corresponding optimized distance $\Delta z_O \sim -15 \: m$ and antenna factor AF($f_O$) $\sim 76.6$ m$^{-1}$, we obtain the value of the ratio between the measured power by the antenna and the total power emitted by the dish given by
\begin{align}
    \gamma_\mathrm{AF} &\approx 3.2 \times 10^{-3} \, .
    \label{ratio_opti}
\end{align}
Note that this order of magnitude agrees with the one reached at the local maximum in Fig.~\ref{fig:3D_plot_gamma}.
The efficiency coefficient obtained for this optimized experimental setup is one order of magnitude larger than the one obtained using the initial set of parameters, Eq.~\eqref{ratio_AF_SHUKET}. Compared to the original 53 loss factor in $\chi$, this optimization leads to a loss of $\sim 17$ on the coupling $\chi$ compared to the initial approximation of $P_\mathrm{rec} = P_\mathrm{dish}$.  
For each frequency of the set $f_\mathrm{set}$, we use the same procedure to obtain a curve of efficiency coefficient as function of the frequency. Then, we apply this frequency-dependent efficiency coefficient to the original constrain on the mixing parameter $\chi$ set by SHUKET \cite{SHUKET}, in order to obtain an estimate of the sensitivity of a new SHUKET-like experiment using these optimized frequencies and distances. However, as can be seen in \cite{SHUKET}, the SHUKET constraint on $\chi$ depends on the frequency, even though the assumed signal is frequency independent. As explained in \cite{SHUKET}, this is due to the frequency dependent gain of the power amplifier connecting the horn antenna to the spectrum analyzer. Since we can conservatively assume that a new run of SHUKET would operate another amplifier with higher gain, we will consider the highest constraint of the original experiment as a basis for our new estimate.

Such estimate is shown by the orange dashed curve in Fig.~\ref{fig:SHUKET_sens_maj}. One can notice that over the $15.5-17.3$ GHz DM frequency range, this new run with optimized parameters would improve the current constraint on the kinetic mixing parameter $\chi$, compared to CMB, shown in blue. 

\section{Discussion and outlook}

In this paper, we have presented an improved modeling of experiments using a dish antenna to search for a DP. We have shown that two effects can possibly impact significantly the expected signal: (i) diffraction effects discussed in Sec.~\ref{sec:propag_field} and (ii) the matching of the mode of the focused electric field with the detector antenna, see Sec.~\ref{sec:detection_field}.

As it was shown in the previous section, the power loss through propagation and detection is highly dependent on the DP frequency one is searching for. For low frequencies experiments, i.e in the GHz range, such as \cite{SHUKET}, the expected sensitivity on the $\chi$ coupling is reduced by a non negligible factor. Indeed, we have shown in Sec.~\ref{sec:SHUKET} that the loss factor which can be estimated using Eq.~\eqref{ratio_AF_SHUKET} implies an overestimation of the constraint on the kinetic mixing parameter $\chi$ by a factor $\sim$ 53.  

Using the calculations performed in this paper, it is possible to  optimize the experimental parameters for this experiment. One consequence of this optimization is to search for DP at higher frequencies.
The optimized experimental parameter leads to an approximate gain of one order of magnitude on the power, as shown in Eq.~\eqref{ratio_opti}, for an optimized frequency $f_O = 16.5$ GHz. The power received in that case is still lower than initially expected in \cite{SHUKET}, but the factor loss in sensitivity is now of order 17.  As it is shown in Fig.~\ref{fig:SHUKET_sens_maj}, one can repeat this procedure for different frequencies, leading to interesting constraints on $\chi$ on a new frequency band. 

Therefore, the detailed optimization scheme could be interesting to use for future experimental runs, in order to be able to constrain this coupling-mass region of the parameter space. Nevertheless, note that the size of the dish antenna used in SHUKET experiment is not optimal for searches at this frequency because diffraction effects can still be significant.

Additional experiments searching for DP at comparable frequencies e.g \cite{Qualiphide, Knirck23, Bajjali23} might also see their constraints revised, due to either diffraction effects or overlap integral of modes. Some further calculations following this paper might be interesting to do to check for such corrections.

Other experiments, such as \cite{Tokyo1, Tokyo3, Tokyo4, FUNK}, considered much higher DP masses, which correspond closely to the optical regime. The loss due to the diffraction in such cases is  insignificant such that we believe that the results from these experiments will not be impacted. 

Note that other papers such as \cite{An23} aim at deriving analytical expressions for electric fields in comparable situations to the present paper. \cite{An23} are mostly interested in experiments using large curvature dish antennas and derive field expressions for any curvature radius, while we focused on low curvature antennas. Nevertheless, one can show that using low curvature approximations on Eq. (S66) of Supplementary Material of \cite{An23}, the detectable power at the curvature center of a dish antenna $(\rho=0)$ is the same as what one would find using Eq.~\eqref{E_field_complete} of the present paper, as expected. Therefore, both methods work and can be complementary.

\begin{acknowledgments}

This work was supported by the Programme National GRAM of CNRS/INSU with INP and IN2P3 cofunded by CNES.

\end{acknowledgments}

\appendix

\section{\label{ap:Green_derivative}Derivation of field emitted by the dish}

We start by the Dirichlet Green function Eq.~\eqref{Dirichlet}. To compute Eq.~\eqref{E_Dirichlet}, we use the directional derivative since $\hat n' = \hat z'$
\begin{subequations}
\begin{align}
&\vec \nabla G_D(\vec{x},\vec{x}')\cdot \hat{n}' = \frac{\partial G_D}{\partial z'}\,\\
&=-\frac{(ikL'-1)(R-a+\Delta z-z')e^{ikL'}}{4\pi L'^3}+\,\nonumber \\
&\frac{(ikL''-1)(R-a-\Delta z-z')e^{ikL''}}{4\pi L''^3}\, ,
\end{align}
\end{subequations}
where
\begin{subequations}
\begin{align}\label{eq:Green_distances}
L' &= \sqrt{\rho^2+\rho'^2-2\rho\rho'\cos(\phi-\phi')+(R-a+\Delta z-z')^2}\,\\
L'' &= \sqrt{\rho^2+\rho'^2-2\rho\rho'\cos(\phi-\phi')+(R-a-\Delta z-z')^2}\, .
    \end{align}
\end{subequations}
Now, plugging the $z'$ coordinate of the plane and considering reception at $\vec x=(\rho,\phi,z)$, we get
\begin{align}
\frac{\partial G_D}{\partial z'}\Big|_\mathrm{z' \ \in \ plane}&= -\frac{(ikL-1)\Delta z}{2\pi L^3}e^{ikL} \, ,
\label{Green_plane}
\end{align}
with $L = \sqrt{\rho^2+\rho'^2-2\rho\rho'\cos(\phi-\phi')+(\Delta z)^2}$.

\section{\label{test_Vinet}Thin optical elements approximation for SHUKET}

In this section, we want to check if Eq.~\eqref{Vinet} is valid in the SHUKET setup, in the case of emission from standing DP electric field. To do so, we propose the following "reverse engineering-like" method : 
\begin{itemize}
\item Consider the field on the dish using boundary conditions Eq.~\eqref{eq:E_out}, i.e $\vec U_D(\vec x) = i\chi \omega \vec Y_{\parallel, D}(\vec x)$,
\item Compute the field at each point $\vec x'$ on the plane closing the dish $\vec U_P(\vec x')$ from Eq.~\eqref{Vinet},
\item Compute numerically the field on the dish $\vec U^\mathrm{test}_D(\vec x)$ from $\vec U_P(\vec x')$ using Kirchhoff integral theorem Eq.~\eqref{E_Dirichlet},
\item Compute the relative error between $\vec U_D(\vec x) $ and $\vec U^\mathrm{test}_D(\vec x)$.
\end{itemize}

We consider the plane located at $z'=R-a$, then the field at a point $(\rho', \phi',z')$ on the plane going towards the dish, located at $z \geq z'$, is given by
\begin{align}
\vec U_\mathrm{P\rightarrow D}(\rho',\phi',z') &= i\chi \omega e^{-ikf(\rho')} \vec Y_{\parallel, D}(\rho',\phi',f(\rho')+z') \, .
\label{E_plane_Vinet}
\end{align}
Notice the change of sign on the wavevector $k$, compared to Eq.~\eqref{Vinet} as we are now considering emission towards the dish (positive $z$ axis). In addition, we aim at computing the field reflected by the dish, while Eq.~\eqref{Vinet} gives the incident field. By boundary conditions, we assume that both incident and reflected are equal, up to a sign, hence the positive sign in front of Eq.\eqref{E_plane_Vinet}.

Using Eqs.~\eqref{E_Dirichlet},\eqref{Green_plane} and \eqref{E_plane_Vinet}, the field in a point on the receiving surface, i.e, the dish, of coordinates $(\rho,\phi,z)$ is given by 
\begin{widetext}
\begin{subequations}
\begin{align}
\vec U^\mathrm{test}_D(\rho,\phi,f(\rho)+R-a) &= -\frac{i\chi \omega f(\rho)}{2\pi}\int_0^r  d\rho' \rho'e^{-ikf(\rho')}\int_0^{2\pi}d\phi' \frac{ikL-1}{L^3}e^{ikL}\vec Y_{\parallel, D}(\rho',\phi', f(\rho')+R-a) \,\\
&\approx -\frac{i\chi \omega f(\rho)}{2\pi}\begin{pmatrix}
Y_x\,\\
Y_y\,\\
0
\end{pmatrix}\int_0^r  d\rho' \rho'e^{-ikf(\rho')}\int_0^{2\pi} d\phi' \frac{ikL-1}{L^3}e^{ikL} \, ,
\end{align}
\end{subequations}
\end{widetext}
where the difference of $z$ positions $\Delta z$ between the fictional plane and the dish from Eq.~\eqref{Green_plane} is now positive and corresponds exactly to $f(\rho)$, $\Delta z=f(\rho) = (r^2-\rho^2)/2R > 0$.
Then, the relative error between $\vec U^\mathrm{test}_D$ and $\vec U_D \approx i \chi \omega \left(Y_x,Y_y,0\right)^T$ (from Eq.~\eqref{field_polarization} is 
\begin{subequations}
\begin{align}\label{error_Vinet}
    &\epsilon(\rho,\phi) \approx \left|\frac{\vec U^\mathrm{test}_D - \vec U^\mathrm{DM}_D}{\vec U^\mathrm{DM}_D}\right|\,\\
    &\approx \left|-\frac{f(\rho)}{2\pi}\left(\int_0^r  d\rho' \rho'e^{-ikf(\rho')}\int_0^{2\pi} d\phi' \frac{ikL-1}{L^3}e^{ikL}\right) - 1 \right| \, .
\end{align}
\end{subequations}
\begin{figure*}
    \centering
    \includegraphics[scale=0.6]{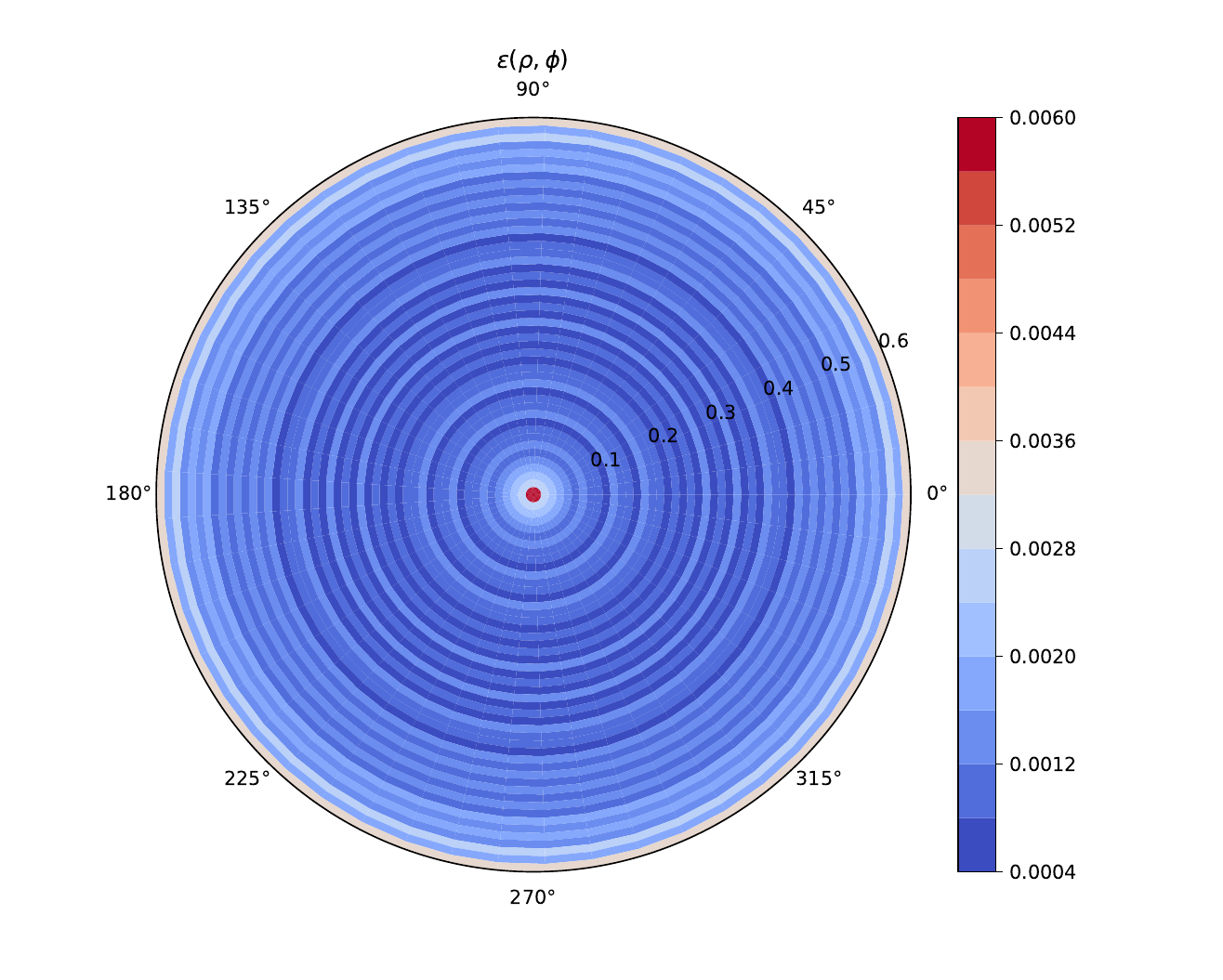}
    \caption{Numerical computation of $\epsilon(\rho,\phi)$ from Eq.~\eqref{error_Vinet} for various dish positions $(\rho,\phi)$. Over the full surface, it is of the order of the \textperthousand.}
    \label{fig:err_relat_Vinet}
\end{figure*}
Fig.~\ref{fig:err_relat_Vinet} shows the numerical computation of this relative error as function of the position $(\rho, \phi)$ on the dish. One can notice that the error is less than 1\%, implying that Eq.~\eqref{Vinet} can be safely used for the propagation of the electric field from the dish in the SHUKET system, with negligible error. Note that the error $\epsilon(\rho,\phi)$ is computed only for points on the dish with radial coordinates $\rho < r$, as the circle of points belonging to the dish with coordinates $(r,\phi,z_\mathrm{plane}), \phi \in [0,2\pi[$ belongs to the fictional plane as well, and Kirchhoff integral is only valid for the computation of the field at reception points outside the emission surface.

\section{\label{FFA_analytic}Analytic expression of the field emitted by SUKET dish}

We start by the electric field Eq.~\eqref{E_field_FFA}, which is trivially obtained from Eq.~\eqref{E_field_complete} by assuming $kL \gg 1$ and Eq.~\eqref{eq:L_low_curv}. We expand the exponential inside the integral of Eq.~\eqref{E_field_FFA}, since the parameter is much smaller than 1. Indeed, for the parameters of SHUKET, we have $\Delta z = -R+a$ (i.e $z_\mathrm{ant}=0$) and 
\begin{align}
    \frac{k\rho'^2}{2}\left(\frac{1}{R}+\frac{1}{\Delta z}\right) &\approx \frac{k \rho'^2 a}{2R^2}< \frac{k r^2 a}{2R^2} = 1.4 \times 10^{-4}\,
\end{align}
Then, the integrand of Eq.~\eqref{E_field_FFA} becomes
\begin{align}
    \rho'(1-i\epsilon \rho'^2)J_0\left(\frac{k\rho\rho'}{|\Delta z|}\right)
\end{align}
at first order in the small parameter $\epsilon \rho'^2$, where $\epsilon=k a/2R^2$. Then, the integral is analytically calculable and the electric field reads 
\begin{subequations}
\begin{align}
    &\vec U^\mathrm{SHUKET}_\mathrm{dish}(\rho,\Delta z) \approx Ae^{i\Phi(\rho)}\frac{r\omega}{\rho}\left(J_1\left(\frac{rk\rho}{|\Delta z|}\right)-\right.\,\nonumber\\
    &\left.i\frac{ar}{2\rho R^2}\left(2\Delta z J_2\left(\frac{rk\rho}{|\Delta z|}\right)-rk\rho J_3\left(\frac{rk\rho}{|\Delta z|}\right)\right)\right) \, \label{E_field_expansion_bessel} ,
\end{align}
where $A=\chi\left(Y_x,Y_y,Y_z\right)^T$. One can verify easily that, with the set of parameters considered, the second term in Eq.~\eqref{E_field_expansion_bessel} containing the second and third Bessel functions is smaller by a factor $\sim 10^{6}$ compared to the other term $\propto$ the first Bessel function. Therefore, we can simplify the expression of the field as
\begin{align}
    &\vec U^\mathrm{SHUKET}_\mathrm{dish}(\rho,\Delta z) \approx \frac{r \omega \chi}{\rho}e^{i\Phi(\rho)}\begin{pmatrix}
Y_x\\
Y_y\\
0
\end{pmatrix}J_1\left(\frac{rk\rho}{|\Delta z|}\right) \, ,
\end{align}
\end{subequations}
and we recover Eq.~\eqref{E_field_final}.

\section{\label{FFA_analytic_opti}Analytic expression of the field emitted by SHUKET dish (optimized parameters)}

We derive an analytic expression of the electric field emitted by the dish in the case of optimized parameters for SHUKET. We will concentrate on one particular optimized frequency $f_O$ of the set $f_\mathrm{set}$, but the same procedure can be applied to the whole set.
We consider the optimized parameters for DM frequency $f_O=16.5$ GHz and dish-antenna distance $\Delta z_O \approx -14.8$ m, the exponent in the integrand of Eq.~\eqref{E_field_FFA} can be comparable to 1, therefore the same Taylor expansion as in Appendix.~\eqref{FFA_analytic} is not possible. Instead, we expand the Bessel function as 
\begin{align}\label{Bessel0_expansion}
    J_0\left(x\right) = \sum_{m=0}^\infty \frac{(-1)^m x^{2m}}{m!4^m\Gamma(m+1)}
\end{align}
In our case, $x=|k_O\rho\rho'/\Delta z_O|$, $k_O = 2\pi f_O/c$. The maximum value of $\rho$ depends on the effective size of the antenna at frequency $\omega_O = 2\pi f_O$ following Eq.~\eqref{eq:effective_surface}. From the antenna gain $G(\omega_O) = 16.87$ dBi, the maximum value of $x$ is $x_\mathrm{max} \approx 0.30$. In the range $[0,x_\mathrm{max}]$, one can easily show that the relative error on $J_0$ by only taking the first two terms of the sum Eq.~\eqref{Bessel0_expansion} is very small $\sim 10^{-4}$, which indicates that these two terms are sufficient to describe the Bessel function in our system. Therefore, the integrand of Eq.~\eqref{E_field_FFA} becomes
\begin{subequations}
\begin{align}
    \rho'e^{-i\varphi(\rho',\Delta z_O)}\left(1-\left(\frac{k_O\rho\rho'}{2\Delta z_O}\right)^2\right)
\end{align}
since
\begin{align}
    \sum_{m=0}^1 \frac{(-1)^m x^{2m}}{m!4^m\Gamma(m+1)} = 1 -\frac{x^2}{4}
\end{align}
\end{subequations}
Then, the analytic integration is doable and gives
\begin{align}
&\vec U^\mathrm{Opti}_\mathrm{dish}(\rho,\Delta z_O) = \frac{A\omega R}{4(\Delta z_O)^2(R+\Delta z_O)^2}e^{ik_0\Delta z_0}\,\\
&\left(2\left(e^{i\Phi'}-1\right)\Delta z_O(k_O\rho^2 R-2i\Delta z_O(R+\Delta z_O))-\right.\,\nonumber \\
&\left.ir^2k^2_O\rho^2(R+\Delta z_O)\right) \nonumber \, ,
\end{align}
with $A=\chi\left(Y_x,Y_y,Y_z\right)^T$ and $\Phi'=k_Or^2(R+\Delta z_O)/2R\Delta z_O$.
From this expression, one can show that the mode associated to this electric field is roughly constant over the effective size of the antenna ($\sim 1\%$ variation), therefore both methods presented in the Sec.~\ref{sec:detection_field} can work to compute the relative power received by the antenna. We find
\begin{subequations}
    \begin{align}
        \gamma_\mathrm{Overlap} &\approx 5.1 \times 10^{-3} \,\\
        \gamma_\mathrm{AF} &\approx 3.2 \times 10^{-3} \, .
    \end{align}
\end{subequations}
Notice that in the same way as for SHUKET parameters, the two results differ by an approximate factor 1.5 difference, as expected. Comparing both values with the ones presented in Sec.~\ref{sec:SHUKET}, one finds an approximate factor 9 improvement in power received. 

\bibliographystyle{apsrev4-1}
\bibliography{DP_shuket} 

\end{document}